# Extreme Raman red shift: ultrafast multimode non-linear space-time dynamics, pulse compression, and broadly tunable frequency conversion

P. A. CARPEGGIANI[1,†,*], G. COCCIA[1], G. FAN[1,2], E. KAKSIS[1], A. PUGŽLYS[1,3], AND A. BALTUŠKA[1,3]

R. PICCOLI[2,†], Y.-G. JEONG[2], A. ROVERE[2], R. MORANDOTTI[2], AND L. RAZZARI[2,*]

B. E. SCHMIDT[4]

A. A. VORONIN[5,6], AND A. M. ZHELTIKOV[5,6,7*]

[1]*Institut für Photonik, Technische Universität Wien, Gußhausstrasse 27/387, 1040 Vienna, Austria*
[2]*Institut National de la Recherche Scientifique (INRS), Centre Énergie, Matériaux et Télécommunications (EMT), Varennes, QC J3X 1S2, Canada*
[3]*Center for Physical Sciences and Technology, Savanoriu Ave. 231, LT-02300, Vilnius, Lithuania*
[4]*few-Cycle Inc., 2890 Rue de Beaurivage, Montreal, H1L 5W5 Quebec, Canada*
[5]*Physics Department, International Laser Center, M. V. Lomonosov Moscow State University, Moscow 119992, Russia*
[6]*Russian Quantum Center, Skolkovo, Moscow Region, 143025 Russia*
[7]*Department of Physics and Astronomy, Texas A&M University, College Station, Texas 77843, USA*
*\*Corresponding authors: paolo.carpeggiani@tuwien.ac.at, razzari@emt.inrs.ca, zheltikov@physics.msu.ru*

*†These authors contributed equally to this work.*

**Ultrashort high-energy pulses at wavelengths longer than 1 μm are nowadays desired for a vast variety of applications in ultrafast and strong-field physics. To date, the main answer to the wavelength tunability for energetic, broadband pulses still relies on optical parametric amplification (OPA), which often requires multiple and complex stages, may feature imperfect beam quality and has limited conversion efficiency into one of the amplified waves. In this work, we present a completely different strategy to realize an energy-efficient and scalable laser frequency shifter. This relies on the continuous red shift provided by stimulated Raman scattering (SRS) over a long propagation distance in nitrogen-filled hollow core fibers (HCF). We show a continuous tunability of the laser wavelength from 1030 nm up to 1730 nm with conversion efficiency higher than 70% and high beam quality. The highly asymmetric spectral broadening, arising from the spatiotemporal nonlinear interplay between high-order modes of the HCF, can be readily employed to generate pulses (~20 fs) significantly shorter than the pump ones (~200 fs) with high beam quality, and the pulse energy can further be scaled up to tens of millijoules. We envision that this technique, coupled with the emerging high-power Yb laser technology, has the potential to answer the increasing demand for energetic multi-TW few-cycle sources tunable in the near-IR.**

## Introduction

An increasing number of applications, such as attosecond pulse isolation via high-harmonic generation (HHG) [1], laser-induced field-driven electron emission [2], laser wakefield acceleration [3], laser-induced electron diffraction [4], optical coherence tomography [5], ultra-broadband [6] and high-field [7] terahertz (THz) generation, were shown to benefit from the increase of the driver pulse wavelength. For instance, the maximum photon energy generated via HHG [1] as well as the energy cut-off of field-driven photo-emitted electrons [2] scale as $\lambda^2$, while the energy of ultrashort THz pulses generated via two-color plasma scales as $\lambda^{4.6}$ [6]. Regarding HHG, longer wavelengths extend the cut-off, but the overall efficiency drops with $\lambda^{-5 \div} \lambda^{-6}$, so each target spectral region has an optimal driver wavelength. For HHG in the water window (280-350 eV), the driver wavelength should be just above 1.2 µm [8]. Therefore, there is a continuously growing interest in generating few-cycle wavelength-tunable IR fields, preferably with a stable carrier-envelope phase. Common laser media based on Ti, Yb, Nd, Cr, Er, and Tm ions, may generate ultrashort pulses, but at specific wavelengths: 800 nm, 1030 nm, 1064 nm, 1350 nm, 1500 nm, and 2000 nm, respectively. To date, the only viable solution providing some degree of frequency tunability relies on optical parametric amplification (OPA) in non-centrosymmetric materials (i.e. second-order nonlinear media featuring $\chi^{(2)}$) for frequency down-conversion [9,10]. Although in typical femtosecond OPA systems the energy conversion efficiency from the pump to the frequency-down-converted wave can reach up to 10% (signal or idler) [11], this approach presents several drawbacks. For instance, an OPA based on BBO or KTA nonlinear optical crystals and pumped by an Yb laser emitting at 1.03 µm can achieve continuous frequency tuning in the near-infrared region >1.35÷2.06 µm for the signal (and 2.06÷<4.5 µm for the idler) with a "gap" in the spectrum between 1.03 µm and 1.35 µm that cannot be filled directly. This is because the transparency of the nonlinear crystal employed, the phase-matching condition, and the temporal matching required to efficiently transfer energy from the pump to the signal/idler set strict boundaries to the operation range. The spectral range between 1.03 µm and 1.35 µm via OPA can be attained either by doubling the pump frequency via second harmonic (SH) generation or by frequency doubling of the generated idler pulses, however, at the expenses of the overall efficiency. Typically, OPAs involve sequential cascaded frequency conversions and require multiple stages, impair beam quality, and deliver pulse durations comparable to the one of the pumping (or seeding) laser system. As an alternative, stimulated Raman scattering (SRS) [12] has firmly established itself as a versatile technique for pulse frequency conversion that enables, among other effects, tunable soliton self-frequency shift in fibers [13], generation of optical attosecond pulses from a collection of Raman sidebands [14], and efficient Stokes-pulse generation in molecular gas cells pumped by narrowband pulses [15]. However, despite numerous attempts, in the case of broadband energetic pulses the generation efficiency of SRS-converted ultrashort pulses does not exceed 15% [16]. A gas-filled HCF is indeed an advantageous platform for extending the non-linear optical interaction length to several meters with low/negligible dispersion. Moreover, because of the waveguiding geometry, the output beam profile has typically a very high quality. In this work we show that, by propagating ~200 fs pulses delivered by Yb-based amplified laser systems (central wavelength of 1030 nm) in a $N_2$-filled HCF (fig. 1a,b), an asymmetric spectral broadening towards longer wavelengths is observed (fig. 1c,d). A fundamental difficulty in increasing the energy and bandwidth of the pump pulses in the SRS scheme is that other nonlinear optical processes, such as self-phase modulation (SPM) and ionization, easily overtake the low-efficiency SRS interaction. Over the past two decades, hollow-core capillary fibers (HCF) have been extensively exploited for extreme pulse compression via nonlinear-induced spectral broadening, using both Raman-inactive noble gases [17-19] and molecular gases [20-22]. Given the bandwidth of the laser systems employed in our investigation (~3.8 THz at $1/e^2$ of the intensity), the interaction of the pulses with the $N_2$ gas is affected only by rotational modes (rotational constant B = 0.06 THz) and is not influenced by the vibrational ones ($\Delta\nu_{vib}$ = 70 THz) [23,24]. The red shift thus results from a cascaded Raman effect evolving during the long propagation (~6 m) in the HCF. By simply adjusting the gas pressure, such system grants the possibility of continuously red-shifting the spectrum by as much as 700 nm (fig. 1d) for the sub-mJ setup (fig. 1b) and by 300 nm (fig. 1c) for the ~10 mJ setup (fig. 1a). We note that the commonly used one-dimensional pulse evolution model [25] cannot explain such impressive spectral asymmetry. Only a full three-dimensional approach, including the spatiotemporal nonlinear dynamics between high-order modes propagating in the HCF, can account for the spectral asymmetry observed in our experiments. In addition to frequency tuning, the obtained spectral broadening permits drastic pulse shortening via post-compression of the whole spectrum (fig. 1a) or via spectral filtering(fig. 1b). This is especially important for Yb-based lasers. Indeed, the latter have been surpassing the well-established Ti:Sapphire systems in terms of overall efficiency and average output power (up to ~10 kW), but are still limited to hundreds of femtosecond (at best) pulse durations [26]. Our strategy can straightforwardly deliver ultrashort pulses with high beam quality, wide frequency tunability and scalable energy. This technique coupled with the emerging high-power femtosecond Yb laser technology, has the potential to satisfy the demand for multi-TW infrared few-cycle sources that are needed in many strong-field applications.

## Experimental setups

The results were produced in two different laboratories at the TUWien and at the INRS-EMT.

~200 fs pulses at 1.03 µm from Yb-based amplified laser systems are coupled into upgraded, soft, stretched HCFs from *few-cycle Inc*. Unlike conventional rigid HCFs, a soft HCF is stretched and kept straight by using specific holders. In this way, HCFs with lengths exceeding several meters can be used to enable high pulse compression ratios at significant, up to 80%, throughput levels. The focal spot size ($1/e^2$) was adjusted to 64% of the fiber core diameter for optimum matching to the fundamental $EH_{11}$ mode, by properly adjusting the distance between two mirrors of a telescope. At the TUWien (fig. 1a), a home-built regenerative amplifier based on a cryogenically cooled Yb:CaF$_2$ delivering up to 14 mJ, 220 fs pulses at 2 kHz was used in combination with a 5.5-m-long, 1 mm inner diameter HCF. A 50 Hz multi-pass Yb:CaF$_2$ laser amplifier was also used for testing input pulse energies up to 26 mJ. The output spectrum was recorded with an optical spectrum analyzer (*Yokogawa AQ-6315A*) while the temporal characterization of the output pulses was performed via second harmonic generation frequency resolved optical gating (SHG-FROG) employing a 50-µm-thick β-BBO crystal. The output pulses were recompressed with 4 bounces on broadband chirped mirrors (*Ultrafast Innovations,* high reflectivity in the range 800÷1200 nm) providing a group delay dispersion (GDD) of -150 fs$^2$ each. At the INRS-EMT (fig. 1b) an Yb:KGW laser (*Pharos, Light Conversion*) delivering up to 1 mJ, 170 fs pulses at 6 kHz was used in combination with a 6-m-long, 530 µm inner diameter HCF. In this case, the output pulses were simply collimated by a concave mirror (focal length of 1 m) and filtered by means of a 4f-compressor setup set at its zero-GDD point. The output spectra were measured combining two spectrometers (*Avantes*) covering the visible (200 -

1000 nm) and the near-infrared (1000 - 1800 nm) regions. The recorded spectra were then corrected by considering the corresponding spectrometer sensitivities. The output pulses were temporally characterized via SHG-FROG (*few-cycle Inc.*) with a 50-μm-thick BBO crystal.

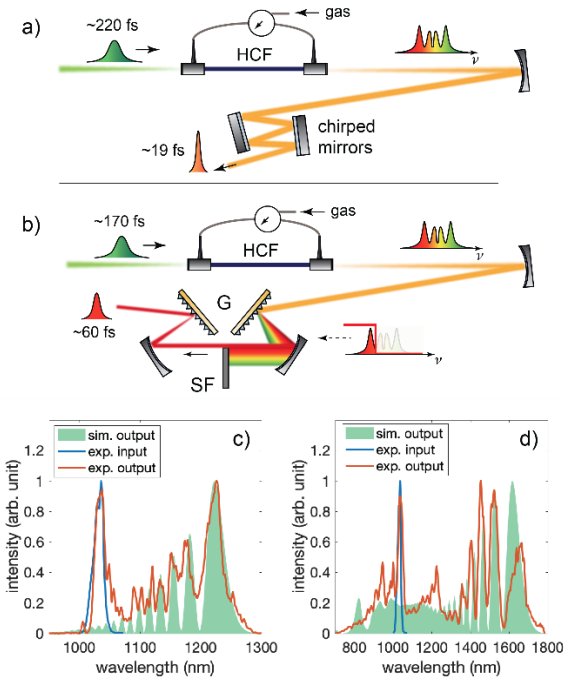

**Fig. 1**. Experimental setup and main result. (a) Experimental setup for ~10 mJ pulses: the laser pulses propagate in a 5.5 m long, 1 mm inner diameter, stretched hollow core fiber (HCF) and are subsequently compressed over the whole bandwidth with chirped mirrors. (b) Experimental setup for sub-mJ pulses: the laser pulses propagate in a 6 m long, 530 μm inner diameter, stretched HCF. The outer lobe on the red side of the spectrum is selected by spatial filtering in a 4f-geometry monochromator. G: grating. SF: spatial filter. (c) Maximum red shift for the setup in (a), obtained at 0.9 bar of $N_2$ pressure. (d) Maximum red shift for the setup in (b), obtained at 4.0 bar of $N_2$ pressure. Experimental data, red line. Simulations, green full curve. Input spectra, blue line. Intensities are normalized.

## Theoretical model

The necessity to develop an adequate 3D propagation model becomes apparent from the shortcomings of the standard 1D model that fails to explain our experimental observations (see SM for detailed information on the theoretical model and physical scenario). Because of the nearly-instantaneous response, lowest-order Kerr nonlinearities cause a symmetric spectral broadening in the case of a symmetric pulse envelope. Correspondingly, the red and blue frequency components are generated at the leading and the trailing pulse edges, respectively. By contrast, the nonlinear phase induced by the Raman effect, due to the delayed response, is shifted towards the trailing edge of the pulse, thus enhancing the red shift. The standard 1D model (fig. 4c in SM), even when it incorporates the Raman-related nonlinearity, predicts significantly more symmetrical broadening for our experimental parameters, thus contradicting our observations. This unusual and in many ways remarkable scenario of extreme spectral red-shifting can be understood considering the complex sequence of strongly-coupled ultrafast multimode spatiotemporal transformations of the ultrashort field waveforms in the gas-filled waveguide.

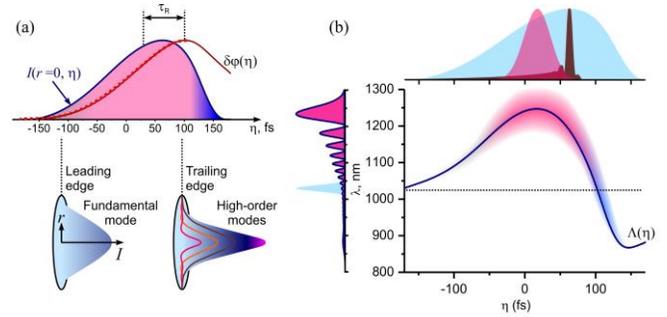

**Fig. 2.** The physics behind extreme Raman red shift. (a) Field intensity envelope (blue solid line) with blue- and red-shifted sections shown by blue and red shading against the temporal phase (red solid line) with its growing part, corresponding to a red spectral shift and represented by red dots. Also depicted are the beam profiles in the leading (left) and trailing (right) edges of the pulse. (b) The instantaneous wavelength Λ as a function of time $\eta$. Shown in the upper panel are the field intensity envelope of the fiber output (light blue), the compressed pulse (brown), and the fiber output transmitted through a 1450 – 1650-nm-bandpass filter (red). Shown on the left panel are the input spectrum (light blue) and the spectrum of the fiber output (red).

Using a quasi-discrete Hankel transformation, the electric field is decomposed into the spatial modes supported by the waveguide (eq. 1 in supplementary materials SM). For the propagation, the delayed Raman response and the instantaneous response due to the Kerr nonlinearity (up to $n_4$) are taken into account (eq. 2 in SM), as well as the dispersion operator, propagation constant and leakage loss of each individual mode. We verified numerically that the ionization term contributes less than 0.1% to the cumulative nonlinear phase shift, dominated by the SPM and the SRS terms, and it is therefore insignificant. Following the treatment of nonlinear propagation in discrete spatial modes, the total electric field is then re-assembled by means of Hankel transformations.

Solving eq. 2 of SM, it is shown that the total nonlinear phase is not uniform both in time (within the pulse envelope) and space (across the beam profile), thus giving rise to a time-dependent nonlinear lens. This lens is responsible for the excitation of higher-order waveguide modes, especially in the trailing edge of the pulse under the combined effect of the delayed Raman response and self-steepening. The walk-off of the modes due to their group velocity mismatch adds to the drastic drop of the field intensity on the trailing edge and increases its lag with respect to the nonlinear phase shift. As a result, the generation of the blue spectral wing is further inhibited (fig. 2a). New red spectral components generated at different times result in a characteristic interference pattern in the spectrum (fig. 2b), visible both in the experiments and in the theory.

The validity of the model is confirmed by the agreement between simulations and experimental results, for both the spectral and temporal properties of the red-shifted pulses.

## Experimental results

### Spectral characterization

For ~10 mJ input pulses in combination with the 1 mm inner diameter fiber, a spectral red shift (defined at the $1/e^2$ level of the peak of the last lobe) up to 1.28 μm was observed, while a shift up to 1.73 μm for ~1 mJ pulses coupled to the 530 μm fiber was measured, as shown in fig. 1c,d and in fig. 3. Up to 82% and 72% of the energy is transferred from the fundamental to longer wavelengths in the first and in the second case, respectively (as calculated by considering the output energy located beyond 1050 nm with respect to the total energy coupled into the waveguide). The different spectral broadenings observed in the two experiments are well understood considering the scaling law introduced in ref. [27], where it was demonstrated that the broadening induced by SPM scales as $\Delta\omega \propto L/A \cdot P \cdot k_2\, p/\lambda_0^2$, where $L$ is the length of the fiber, $A$ the section area, $P$ is the pulse peak power, $\lambda_0$ is the central wavelength and $k_2$ is the ratio between the nonlinear index coefficient and the gas pressure $p$. Even though in our case SRS is the dominant process, we can assume – in a first approximation – the same dependence. The results of the two experimental setups, obtained with:
i) 10 mJ, 220 fs, 0.9 bar, 1 mm inner diameter, and 5.5 m HCF length and
ii) 1 mJ, 170 fs, 4 bar, 530 μm inner diameter, 6 m HCF length
can be compared with this scaling law to obtain a ratio of 1:2, which well corresponds to the experimentally obtained 57 THz and 115 THz red shifts, respectively. As in the case for SPM, also for SRS the maximum broadening is limited by the ionization of the gas and by the critical power for self-focusing, $P_{cr} = \lambda_0^2/(2 k_2 p)$.

The pulse energy $W$ scaling requires larger $A \propto W$ to keep the laser intensity below the limit of ionization of the gas. The limit of the critical power requires to keep constant the product $P \cdot p$, thus $L$ should scale as well as $L \propto A \propto W$ in order to achieve the same $\Delta\omega$. Given the limits of a setup in terms of length, these conditions imply that for (SPM) SRS there is a trade-off between the maximum (broadening) shift and the pulse energy.

When comparing the experimental spectra and the simulations, the only noticeable difference is the persistence of the fundamental wavelength in the data from both setups. This might be due to its propagation in higher-order modes, above the simulation limit, thus resulting in different output spatial distributions. It will be shown in the next paragraph and in the SM that the fundamental wavelength does not affect non-linear applications such as SHG-FROG.

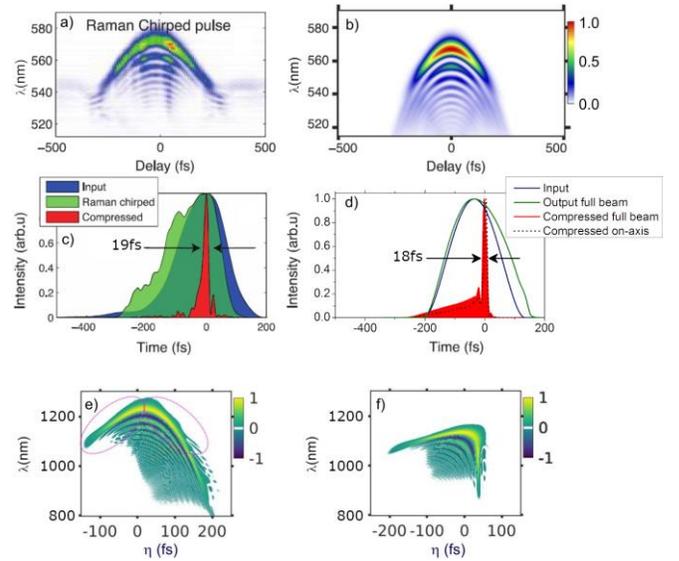

**Fig. 4.** Compression of pulses after the hollow waveguide with an inner diameter $d$ = 1 mm and length $L$ = 5.5 m, filled with N$_2$ at a pressure $p$ = 0.6 bar, for an input pulse with an energy $W$ = 6.5 mJ and a width $\tau$ = 220 fs: (a, b) experimental (a) and simulated (b) SHG-FROG traces of the pulse at the output of the fiber, (c) temporal envelopes of the input (blue shading), output (green shading) and compressed (red shading) pulses retrieved from the experimental SHG-FROG traces, (d) simulated temporal envelopes of the pulse at the fiber output integrated over the beam (green line) and compressed pulse on the beam axis (dashed line) and integrated over the beam profile (red shading) against the input pulse envelope used in the model (blue line). (e, f) Simulated Wigner functions for (e) the pulse at the fiber output and (f) the compressed pulse. Interfering field components are contoured by dotted lines.

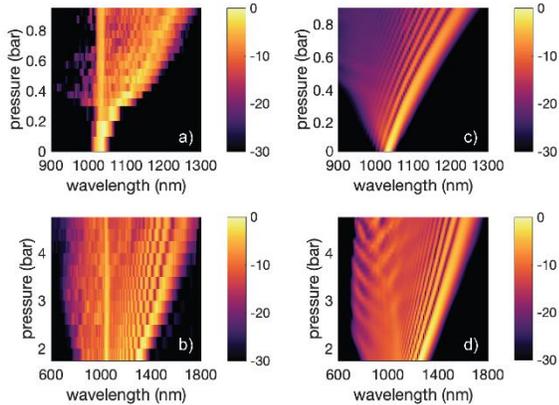

**Fig. 3.** Spectra of the fiber output measured (a, b) and calculated (c, d) as a function of the N$_2$ pressure inside the fiber for (a, c) $W$ = 10 mJ, $\tau$ = 220 fs, $d$ = 1.0 mm and $L$ = 5.5 m and for (b, d) $W$ = 1.0 mJ, $\tau$ = 170 fs, $d$ = 0.53 mm and $L$ = 6.0 m.

**Temporal characterization**

Two possible techniques have been considered for the production of compressed pulses:
i) compressing the whole spectrum with broadband chirped mirrors (fig.1a) and
ii) spectrally selecting the external lobe of the red-shifted spectrum by spatial filtering with a 4f setup set at the zero-GDD point (fig.1b).
The temporal characterizations of the SRS red-shifted pulses for the two experimental setups are shown in fig.4 and fig.5, respectively.
The first approach is suitable for HHG applications, where it is important to minimize the energy loss and preserve the full bandwidth of the red-shifted spectrum. In this case, compressibility was confirmed experimentally (fig. 4c) by exploiting the partial overlap (from 1030 to 1170 nm) in the bandwidth between the red-shifted pulses and the broadband chirped mirrors used for pulse compression, in turn leading to the generation of 19 fs, 6 mJ pulses centered at 1120 nm.

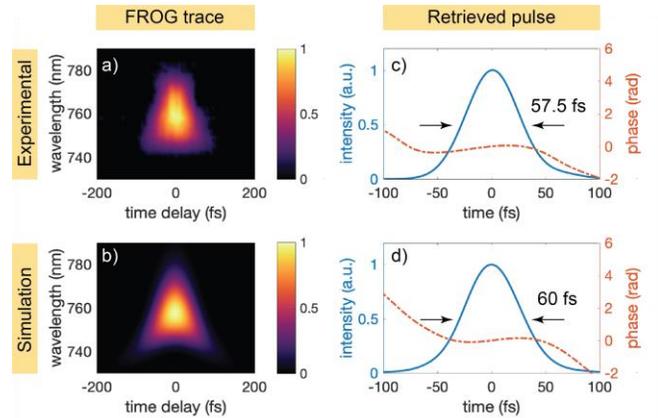

**Fig. 5.** Experimental (a) and simulated (b) SHG-FROG traces of the fiber output transmitted through a 1450 – 1650-nm-bandpass filter. Input pulse with energy $W$ = 0.9 mJ and pulse-width $\tau$ = 170 fs transmitted through a hollow fiber with an inner diameter d = 0.53 mm and length L = 6.0 m, filled with N2 at a pressure p = 3.4 bar. (c, d) Temporal envelope (solid line) and temporal phase (dotted line) retrieved from the experimental (c) and simulated (d) SHG-FROG traces.

Further tests at higher energy (26 mJ) with lower repetition rate allowed us to get an output of 20 mJ (76%) from the fiber, of which 14 mJ (70%) were compressed below 20 fs (corresponding to 0.7 TW peak power). With the second approach, quasi Fourier limited pulses can be directly generated in the near-IR spectral range. This is possible since, analogously to the case of pure SPM [25], the spectral phase over the external lobes assumes a quasi linear dependence over frequency, which in turn results in a simple temporal shift of the corresponding pulse. As an example, fig. 5 shows for the filtered external lobe (FWHM = 73 nm centered at 1510 nm) reached at 3.4 bar $N_2$ pressure, the experimental and theoretical SHG-FROG traces and the corresponding retrieval. The well-behaved spectral modulation achievable with $N_2$ allows us to straightforwardly obtain well-shaped ~57 fs (FWHM – time-bandwidth product $\cong$ 0.55) pulses by simply filtering out the external lobe (~8% of the input energy). This approach guarantees full tunability and is also suitable to access the mid-IR domain with stable CEP, by selecting two lobes of the red-shifted spectrum and performing difference-frequency generation (DFG) in a nonlinear crystal. DFG tuning can be simply achieved by adjusting the phase matching and the time delay between the pulse and its red-shifted replica.

## Conclusions

In conclusion, we have demonstrated a highly attractive energy-scalable approach to develop wavelength-tunable drivers for strong-field applications. Over 70% of the energy of 200-fs, 1030-nm pulses can be transferred into the near-IR spectral domain extending up to 1.73 μm. This breakthrough is enabled by the use of large-core-diameter, multi-meter-long HCFs that, in combination with relatively long input laser pulses and Raman-active gases, provide conditions to promote a distinctive cascaded SRS process over other nonlinear interactions such as SPM. We have shown that the Raman-shifted output spectrum can be compressed with broadband chirped mirrors to obtain few-cycle pulses in the near-IR, or simply filtered to obtain wavelength-tunable ~60-fs-long pulses. We believe that this technique offers a convenient alternative to the more demanding OPA systems to achieve a gap-free tuning in the near-IR, while it may also grant access to the visible domain by SH, as well as to the mid-IR by performing DFG between different lobes of the output spectrum. Moreover, such approach can be up-scaled into the TW regime with the current setup, and an even shorter pulse duration is expected with chirped mirrors covering the entire available spectrum. We envision that the proposed method, coupled with the emerging high-power Yb laser technology, has the potential to answer the increasing demand for long-wavelength multi-TW few-cycle sources needed in many strong-field applications.

**Funding**. Natural Sciences and Engineering Research Council of Canada (NSERC) (Collaborative Research and Development, Strategic, and Discovery Grants); PROMPT.
Austrian Science Fund (FWF) Project # P27491.
Russian Foundation for Basic Research (18-02-40034, 18-32-20196, 18-29-20031, 19-02-00473); Welch Foundation (Grant No. A-1801-20180324); Russian Science Foundation (project No. 20-12-00088 -- multioctave nonlinear optics).
The authors declare no conflicts of interest.

**Acknowledgment**. We wish to thank Prof. Marco Marangoni from Politecnico di Milano for the useful discussion.

See Supplement 1 for supporting content.